\documentclass[aps,prl,twocolumn,showpacs,preprintnumbers,nofootinbib,amsmath,amssymb]{revtex4}

\usepackage{graphicx}
\usepackage{dcolumn}
\usepackage{bm}
\usepackage{amsmath}

\newcommand{\beq}{\begin{eqnarray}}
\newcommand{\eeq}{\end{eqnarray}}

\newcommand{\real}{{\sf I}\kern-.12em{\sf R}}
\newcommand{\comp}{{\sf I}\kern-.50em{\sf C}}
\newcommand{\unity}{{\sf I}\kern-.54em{\sf 1}}

\def\spose#1{\hbox to 0pt{#1\hss}}
\def\ltapprox{\mathrel{\spose{\lower 3pt\hbox{$\mathchar"218$}}
 \raise 2.0pt\hbox{$\mathchar"13C$}}}

\begin{document}

\title{$\theta$-dependence of the deconfinement temperature in Yang-Mills theories.}
\author{Massimo D'Elia}
\affiliation{
Dipartimento di Fisica dell'Universit\`a
di Pisa and INFN - Sezione di Pisa,\\ Largo Pontecorvo 3, I-56127 Pisa, Italy}
\email{delia@df.unipi.it}
\author{Francesco Negro}
\affiliation{Dipartimento di Fisica dell'Universit\`a
di Genova and INFN - Sezione di Genova,\\
 Via Dodecaneso 33, I-16146 Genova, Italy}
\email{fnegro@ge.infn.it}
\date{\today}

\begin{abstract}
We determine the $\theta$ dependence of the deconfinement temperature of 
SU(3) pure gauge theory, finding
that it decreases in presence of a topological $\theta$ term.
We do that by performing lattice simulations at imaginary 
$\theta$, then exploiting analytic continuation.
We also give an estimate of such dependence
in the limit of a large number of colors $N$,
and compare it with our numerical results.
\end{abstract}

\maketitle

The possible effects of a CP violating term in Quantum ChromoDynamics (QCD)
have been studied since long.
Such term enters
the Euclidean lagrangian as follows:
\beq
{\cal L}_\theta  &=& {\cal L}_{\rm QCD} 
- i \theta q(x) \nonumber \\
q(x) &=& \frac{g_0^2}{64\pi^2} \epsilon_{\mu\nu\rho\sigma} F_{\mu\nu}^a(x)
F_{\rho\sigma}^a(x)
\label{thetaterm}
\eeq
where $q(x)$ is the topological charge density.

Experimental upper bounds on $\theta$ are quite stringent
($|\theta| \lesssim 10^{-10}$), suggesting
that such term may be forbidden by some mechanism. 
Nevertheless, the dependence of QCD and of $SU(N)$ gauge theories
on $\theta$ is of great theoretical and phenomenological interest.
$\theta$ derivatives of the vacuum free energy, computed 
at $\theta = 0$, enter various aspects of hadron phenomenology;
an example is the topological susceptibility 
$\chi \equiv \langle Q^2 \rangle/V$ 
($Q \equiv \int d^4 x\ q(x)$ and $V$ is the space-time volume) 
which enters the solution of the so-called $U(1)_A$ problem~\cite{u1wit,u1ven}.
Moreover it has been proposed~\cite{kharzeev} that topological
charge fluctuations may play an important role at finite temperature 
$T$, especially around the deconfinement transition, where local
effective variations of $\theta$ may be detectable
as event by event $P$ and $CP$ violations in heavy ion collisions.

In the present work we study the effect of 
a non-zero $\theta$ on the critical deconfining temperature $T_c$,
considering the case of 
pure Yang-Mills theories. Due to the symmetry under CP 
at $\theta = 0$, the critical temperature $T_c (\theta)$ is
expected, similarly to the free energy, to be an even function
of $\theta$. Therefore we parameterize $T_c (\theta)$ as follows
\beq
\frac{T_c(\theta)}{T_c(0)} = 1 - R_\theta\ \theta^2 + O(\theta^4)
\label{Tcdep}
\eeq

In the following we shall determine $R_\theta$
for the $SU(3)$ pure gauge theory, obtaining $R_\theta > 0$, 
and compare it with a simple model
computation valid in the large $N$ limit, showing that
$R_\theta$ is expected to 
be $O(1/N^2)$.

{\it The method} -- Effects related to the topological $\theta$ term are typically
of non-perturbative nature, hence numerical simulations on a lattice
represent the ideal tool to explore them. However, it is well
known that the Euclidean path integral representation of the partition
function
\begin{equation}
Z(T,\theta) = 
\int [dA]\ e^{ -S_{QCD} [A] + i \theta Q[A]}
=e^{ - {V_s f(\theta)}/{T}},
\label{partfun}
\end{equation}
is not suitable for Monte-Carlo simulations because the measure is complex
when $\theta \neq 0$. 
$S_{QCD} = \int d^4 x\ {\cal L}_{QCD}$
and periodic boundary conditions are assumed over the compactified
time dimension of extension $1/T$;
$f(\theta)$ is the free energy density
and $V_s$ is the spatial volume.

A similar sign problem is met for QCD at finite baryon chemical 
potential $\mu_B$, 
where the fermion determinant becomes complex. In that case, 
a possible partial solution is to study the theory at 
imaginary $\mu_B$, where the sign problem disappears, 
and then make use of analytic continuation to infer the dependence at
real $\mu_B$, at least for small values of $\mu_B/T$~\cite{immu}. 
An analogous approach has been proposed for exploring
a non-zero $\theta$~\cite{azcoiti,alles_1,aoki_1,vicari_im};
as for  $\mu_B \neq 0$, also in this case one
assumes that the theory is analytic around $\theta = 0$,
a fact supported by our present knowledge about 
free energy derivatives at $\theta = 0$~\cite{vicari_rep,alles_2}.

Various studies have shown that 
the dependence of the 
critical temperature on the baryon chemical potential, 
$T_c(\mu_B)$, can be determined reliably up to the quadratic order in 
$\mu_B$, while ambiguities related to the procedure of analytic
continuation may affect higher order terms~\cite{immu_cea}.
It is natural to assume that a similar scenario takes place for 
analytic continuation from an imaginary $\theta \equiv i \theta_I$ term, i.e. that 
$R_\theta$ can be determined reliably from numerical studies
of the lattice partition function:
\begin{equation}
Z_L(T,\theta) = 
\int [dU]\ e^{ -S_L [U] - \theta_L Q_L[U]} \, ,
\label{partfunlat}
\end{equation}
where $[dU]$ is the integration over the elementary gauge link variables
$U_\mu$; $S_L$ and $Q_L$ are the 
lattice discretizations of respectively the 
pure gauge action and the topological charge, 
$Q_L = \sum_x q_L(x)$. We will consider the Wilson
action,   
$S_L = \beta \sum_{x,\mu>\nu} (1 - {\rm Re} {\rm Tr}\, 
\Pi_{\mu\nu}(x)/N)$ where $\beta={2N / g_0^2}$ and 
$\Pi_{\mu\nu}$ is the plaquette operator.

Various choices are possible for the lattice 
operator $q_L(x)$, which in general are linked to the continuum $q(x)$ 
by a finite multiplicative renormalization~\cite{zetaref}
\beq
 q_L(x) {\buildrel {a \rightarrow 0} \over \sim} a^4 Z(\beta) q(x) + O(a^6) \; ,
\label{eq}
\eeq
where $a = a(\beta)$ is the lattice spacing and $\lim_{a \to 0} Z = 1$. 
Hence, as the continuum limit is approached, 
the imaginary part of $\theta$ is related to the lattice parameter
$\theta_L$ appearing in Eq.~(\ref{partfunlat}) 
as follows:
$\theta_I = Z\, \theta_L$.

Since $q_L(x)$
enters directly the functional integral measure, it is important, in order to 
keep
the Monte-Carlo algorithm efficient enough, to choose a simple definition,
even if the associated renormalization is large. Therefore, following
Ref.~\cite{vicari_im}, we adopt the gluonic definition 
\beq
q_L(x) = {{-1} \over {2^9 \pi^2}} 
\sum_{\mu\nu\rho\sigma = \pm 1}^{\pm 4} 
{\tilde{\epsilon}}_{\mu\nu\rho\sigma} \hbox{Tr} \left( 
\Pi_{\mu\nu}(x) \Pi_{\rho\sigma}(x) \right) \; ,
\label{eq:qlattice}
\eeq
where ${\tilde{\epsilon}}_{\mu\nu\rho\sigma} = {{\epsilon}}_{\mu\nu\rho\sigma}
$ for positive directions  and ${\tilde{\epsilon}}_{\mu\nu\rho\sigma} =
- {\tilde{\epsilon}}_{(-\mu)\nu\rho\sigma}$. With this choice gauge links
still appear linearly in the modified action, hence a standard heat-bath 
algorithm over $SU(2)$ subgroups, combined with over-relaxation, can be 
implemented.

Finite temperature $SU(N)$ pure gauge theories possess the so-called
center symmetry,
corresponding to a multiplication of all
parallel transports at a fixed time by an element of the center
$Z_N$. Such
symmetry is spontaneously broken at the deconfinement transition and
the Polyakov loop is a suitable order parameter.
Since $q_L (x)$ is a sum over closed local loops,
the modified action $S_L + \theta_L Q_L$ is also center symmetric, hence
we still expect $Z_N$ spontaneous breaking and we will adopt 
the Polyakov loop and its susceptibility as probes for deconfinement
\beq
\langle L \rangle 
&\equiv& \frac{1}{V_s} \sum_{\vec{x}} \frac{1}{N} \langle {\rm Tr}\
       \prod_{t=1}^{N_t} U_0({\vec x},t) \rangle \, \nonumber \\
\chi_L &\equiv& V_s\ 
(\langle L^2 \rangle - \langle L \rangle^2 )
\rangle \, , 
\label{obs}
\eeq
where $N_t$ is the number of sites in the temporal direction.

\begin{table}
\begin{center}
\begin{tabular}{|c|c|c|c|c|}
\hline
lattice & $\theta_L$ & $\beta_c$ & $\theta_I$ & $T_c(\theta_I)/T_c(0)$ \\ \hline
$16^3 \times 4\ $  &   0\ & 5.6911(4)\ & 0\ & 1 \\
\hline
$16^3 \times 4\ $  &   5\ & 5.6934(6)\ & 0.370(10)\ & 1.0049(11)\ \\
\hline
$16^3 \times 4\ $  &  10\ & 5.6990(7)\ & 0.747(15)\ & 1.0171(12)\ \\
\hline
$16^3 \times 4\ $  &  15\ & 5.7092(7)\ & 1.141(20)\ & 1.0395(11)\ \\
\hline
$16^3 \times 4\ $  &  20\ & 5.7248(6)\ & 1.566(30)\ & 1.0746(10)\ \\
\hline
$16^3 \times 4\ $  &  25\ & 5.7447(7)\ & 2.035(30)\ & 1.1209(10)\ \\
\hline
\hline
$24^3 \times 6\ $  &  0\   & 5.8929(8)\   & 0\ & 1\ \\
\hline
$24^3 \times 6\ $  &  5\   & 5.8985(10)\ & 0.5705(60)\ & 1.0105(24)\ \\
\hline
$24^3 \times 6\ $  &  10\ & 5.9105(5)\  & 1.168(12)\ & 1.0335(18)\ \\
\hline
$24^3 \times 6\ $  &  15\ & 5.9364(8)\   & 1.836(18)\ & 1.0834(23)\ \\
\hline
$24^3 \times 6\ $  &  20\ & 5.9717(8)\   &  2.600(24)\ & 1.1534(24)\ \\
\hline
\hline
$32^3 \times 8\ $  &   0\ & 6.0622(6)\   & 0\ & 1\  \\
\hline
$32^3 \times 8\ $  &   5\ & 6.0684(3)\   & 0.753(8)\ &  1.0100(11)\     \\
\hline
$32^3 \times 8\ $  &   8\ & 6.0813(6)\   & 1.224(15)\ &  1.0312(14)\     \\
\hline
$32^3 \times 8\ $  &   10\ & 6.0935(11)\ & 1.551(20)\ &  1.0515(21)\  \\
\hline
$32^3 \times 8\ $  &   12\ & 6.1059(21)\ & 1.890(24)\ & 1.0719(34) \\
\hline
$32^3 \times 8\ $  &   15\ & 6.1332(7)\   & 2.437(30)\ & 1.1201(17)\ \\
\hline
\end{tabular}
\end{center}
\caption{Collection of results obtained for $\beta_c$ and $T_c$.}
\label{tab:res}
\end{table}

\begin{figure}[h!]
\includegraphics*[width=0.47\textwidth]{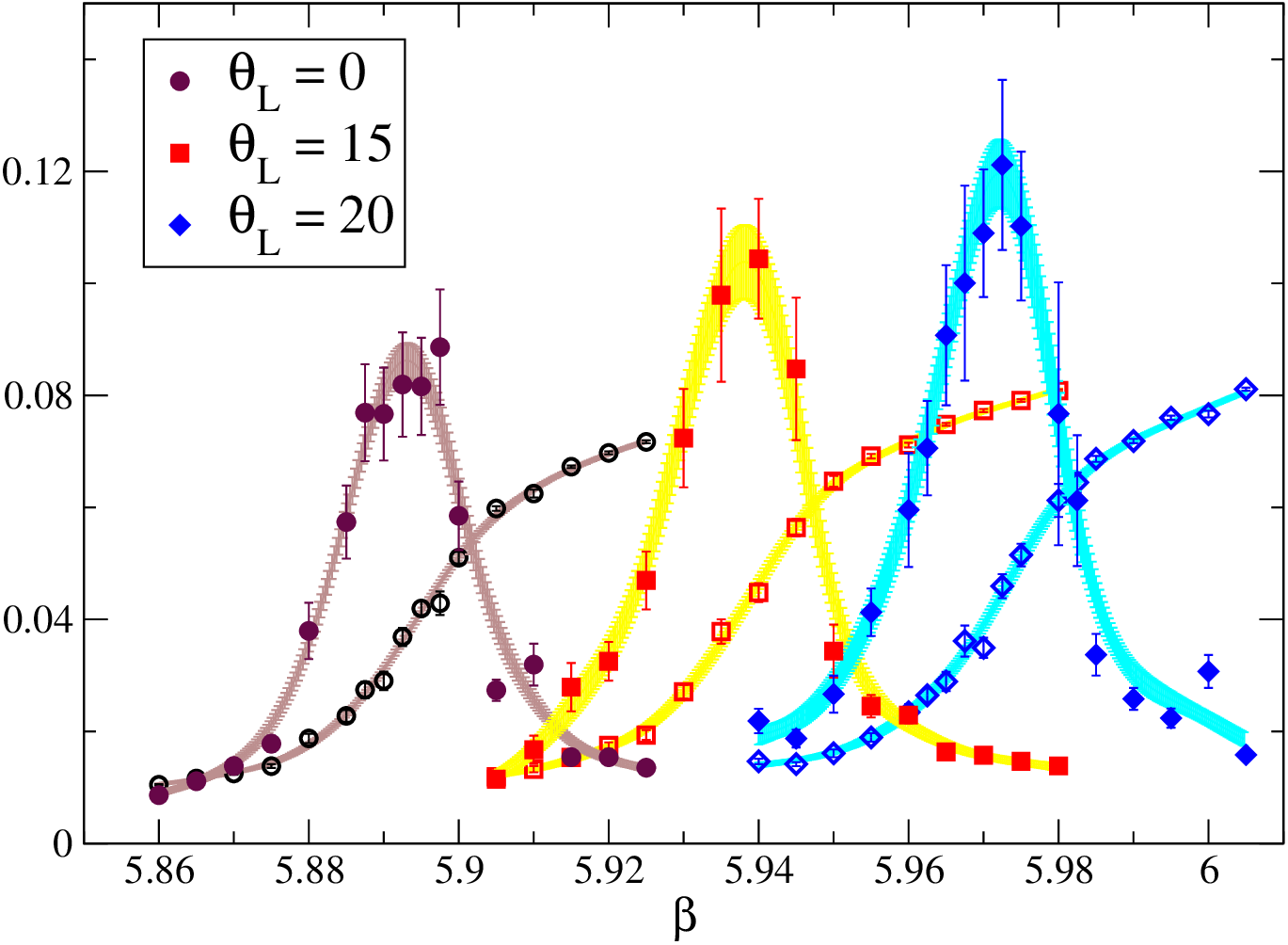}
\caption{Polyakov loop and its susceptibility as a function of 
$\beta$ on a $24^3 \times 6$ lattice and for a few $\theta_L$ values.
The susceptibility values have been multiplied by a factor 250.}
\label{fig:pol}
\end{figure}

{\it Results} -- In the following we present results obtained on three different
lattices, $16^3 \times 4$, $24^3 \times 6$ and $32^3 \times 8$,
corresponding, around $T_c$, to equal spatial volumes
(in physical units) and three different lattice spacings 
$a \simeq 1/(4 T_c)$, 
$a \simeq 1/(6 T_c)$ and 
$a \simeq 1/(8 T_c)$. That will permit us to
extrapolate
$R_\theta$ to the continuum limit.

We have performed, on each lattice, several series of simulations
at fixed $\theta_L$ and variable $\beta$. Typical statistics have
been of $10^5-10^6$ measurements, each separated by a cycle of
4 over-relaxation + 1 heat-bath sweeps, for each run; autocorrelation 
lengths have gone up to $O(10^3)$ cycles around the transition.  
In Fig.~\ref{fig:pol} we show results for the Polyakov loop modulus
and its susceptibility as a function of $\beta$ 
for a few values of $\theta_L$ on a $24^3 \times 6$ lattice;
we also show data obtained after reweighting in $\beta$.
We notice a slight increase in the height of the susceptibility
peak as $\theta_L$ increases, however any conclusion regarding the 
influence of $\theta$ on the strength of the transition would require
a finite size scaling analysis and is left to future studies.

The critical coupling $\beta_c(\theta_L)$ is located at the maximum
of the susceptibility through a Lorentzian fit to 
unreweighted data: values obtained at $\theta_L = 0$ coincides within
errors with those found in previous works~\cite{karsch_thermo}.
From $\beta_c(\theta_L)$ we reconstruct
$T_c(\theta_L)/T_c(0) = a(\beta_c(0))/a(\beta_c(\theta_L))$ by means of 
the non-perturbative determination of $a(\beta)$ reported in 
Ref.~\cite{karsch_thermo}. 
Notice that most finite size effects in the 
determination of $\beta_c(\theta_L)$ should cancel when computing the
ratio $T_c(\theta_L)/T_c(0)$.
A complete set of results is reported
in Table~\ref{tab:res}.

\begin{figure}[h!]
\includegraphics*[width=0.47\textwidth]{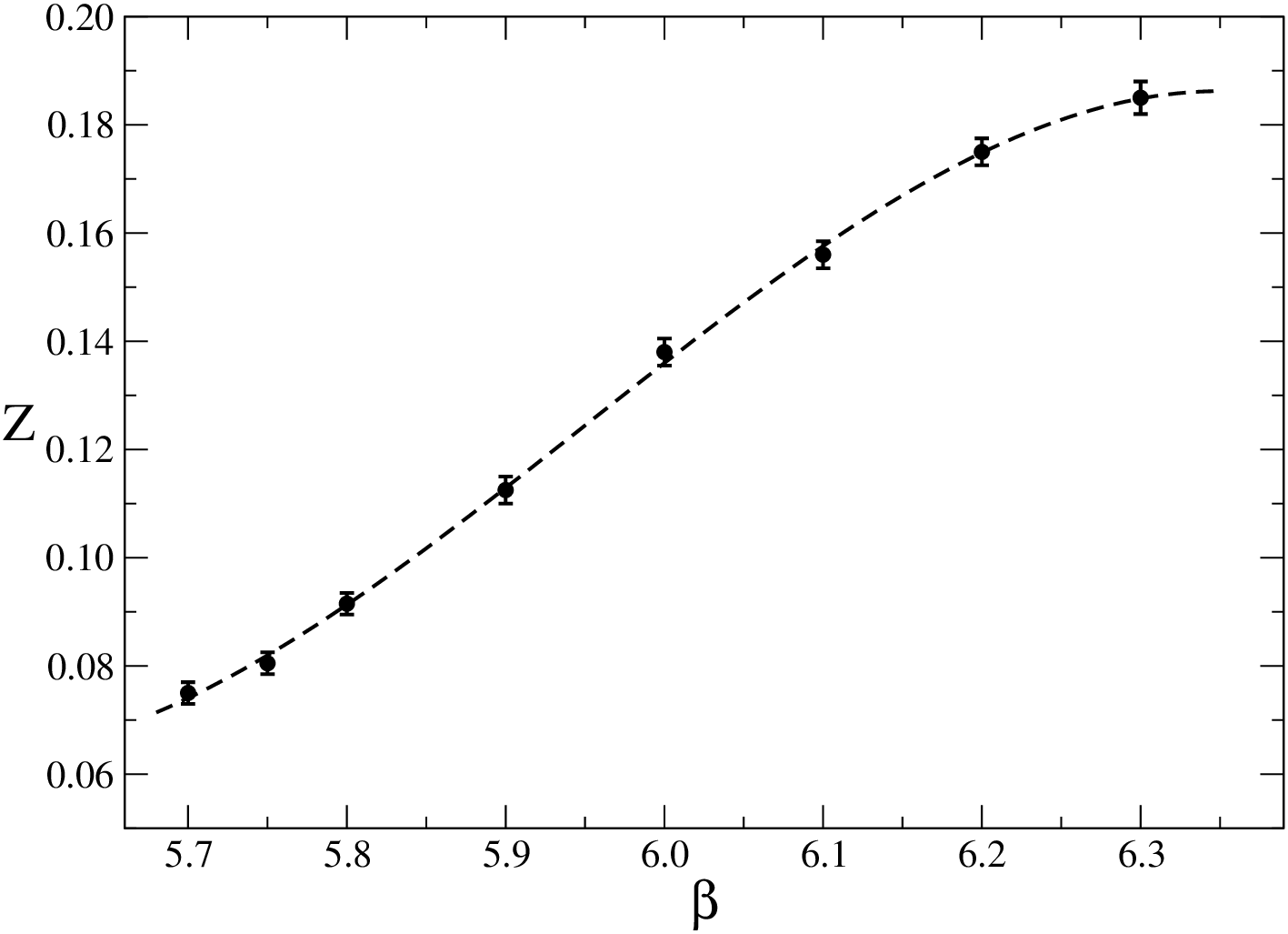}
\caption{Determinations of the renormalization constant $Z$ on a $16^4$ lattice. The 
dashed line is a cubic interpolation of data.}
\label{fig:zeta}
\end{figure}

As a final step, we need to convert $\theta_L$ into the physical
parameter $\theta = i\, \theta_I$. A well known method 
for a non-perturbative determination
of the renormalization constant $Z = Z(\beta)$ is that based
on heating techniques~\cite{ref:heating}. Here we follow
the method proposed in Ref.~\cite{vicari_im}, giving $Z$ in terms
of averages over the thermal ensemble:
\beq
Z = \frac{\langle Q Q_L \rangle}{\langle Q^2 \rangle}
\eeq
where $Q$ is, configuration by configuration, the integer closest to
the topological charge obtained after cooling. Such method assumes, as usual,
that UV fluctuations responsible for renormalization are independent 
of the topological background.
$Z$ has been determined for a set of $\beta$ values on a symmetric
$16^4$ lattice, as reported in Fig.~\ref{fig:zeta},
then obtaining $Z$ at the critical values of $\beta$ by a cubic 
interpolation. Typical statistics have been of $10^5$ measurements, each separated by 5 cycles of
4 over-relaxation + 1 heat-bath sweeps, for each $\beta$; the autocorrelation
length of $Q$ has reached a maximum of $10^3$ cycles at the highest value
of $\beta$.  
A check for systematic
effects has been done by repeating the determination with a different number of cooling sweeps
to obtain $Q$ (15, 30, 45 and 60) 
or, at the highest explored value of $\beta$, 
on a larger $24^4$ lattice.
In this way we finally obtain 
$\theta_I (\beta_c(\theta_L)) = Z(\beta_c(\theta_L))\, \theta_L$, as reported
in the 4th column of Table~\ref{tab:res}.

\begin{figure}[h!]
\includegraphics*[width=0.47\textwidth]{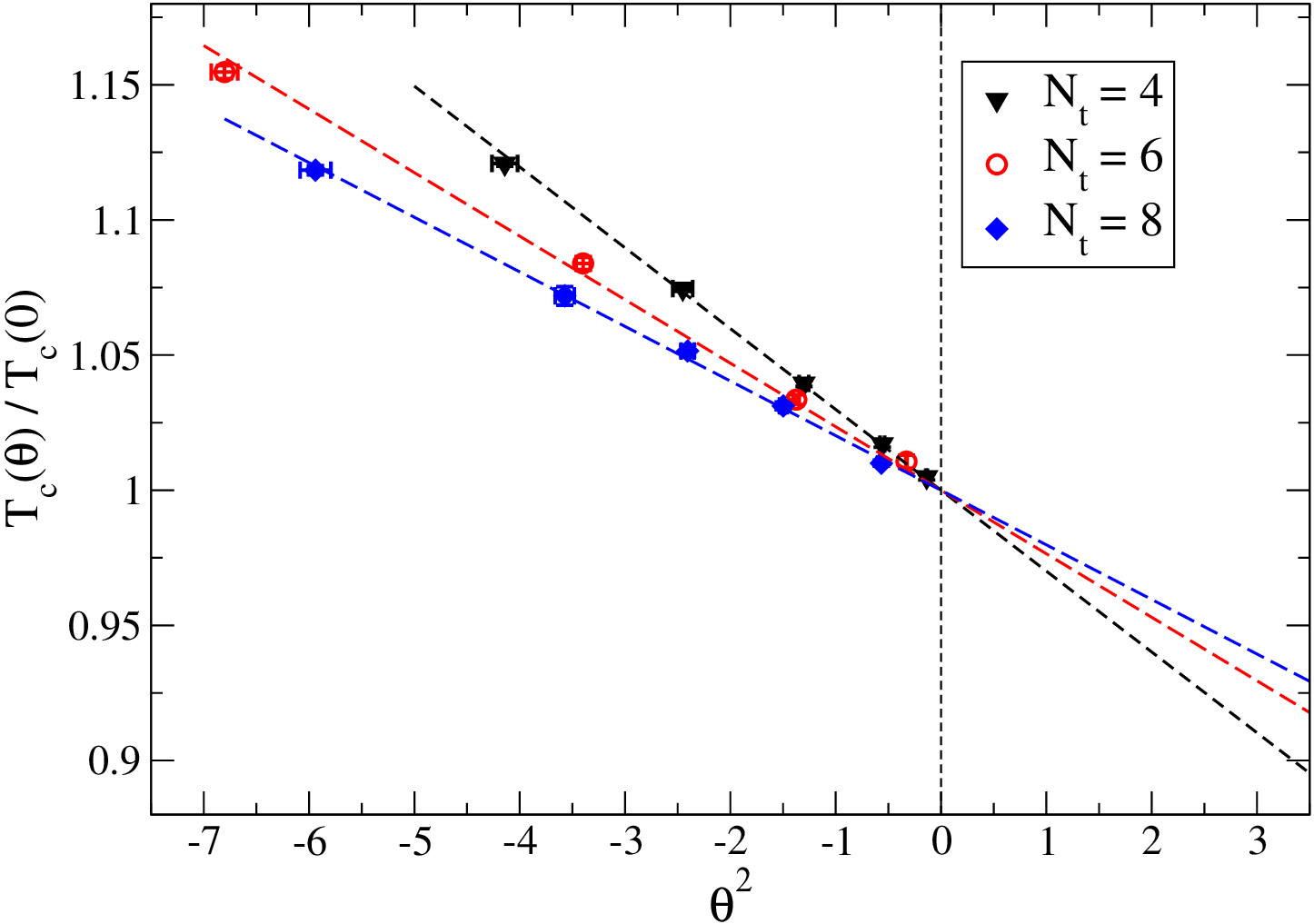}
\caption{$T_c(\theta)/T_c(0)$ as a function of $\theta^2$ for different 
values of $N_t$. Dashed lines are the result of linear fits, as reported in the text,
then extrapolated to $\theta^2 > 0$.}
\label{fig:tctheta}
\end{figure}

Final results for $T_c(\theta_I)/T_c(0)$ and for the three different
lattices explored are reported in Fig.~\ref{fig:tctheta}. In all 
cases a linear dependence in $\theta^2$, according to Eq.~(\ref{Tcdep}), 
nicely fits data. In particular we obtain
$R_\theta = 0.0299(7)$ for $N_t = 4$
($\chi^2/{\rm d.o.f.} \simeq 0.3$), 
$R_\theta = 0.0235(5)$ for $N_t = 6$
($\chi^2/{\rm d.o.f.} \simeq 1.6$) and 
$R_\theta = 0.0204(5)$ for $N_t = 8$
($\chi^2/{\rm d.o.f.} \simeq 0.7$). 

We have performed various tests to check the stability of our fits.
If we change the fit range, e.g. by excluding, for each $N_t$, the
1-2 largest values of $\theta_I$,
results for $R_\theta$ are stable within errors. If we assume a generic 
power like behavior $T_c(\theta)/T_c(0) - 1 = A\ \theta^\alpha$, we always
obtain that $\alpha$ is compatible with 2 within errors; if we fix 
$\alpha$ to values which would imply a non-analyticity at $\theta = 0$,
e.g. $\alpha = 1$, we obtain a $\chi^2/{\rm d.o.f.}$ of $O(10)$ or larger.

\begin{figure}[h!]
\includegraphics*[width=0.47\textwidth]{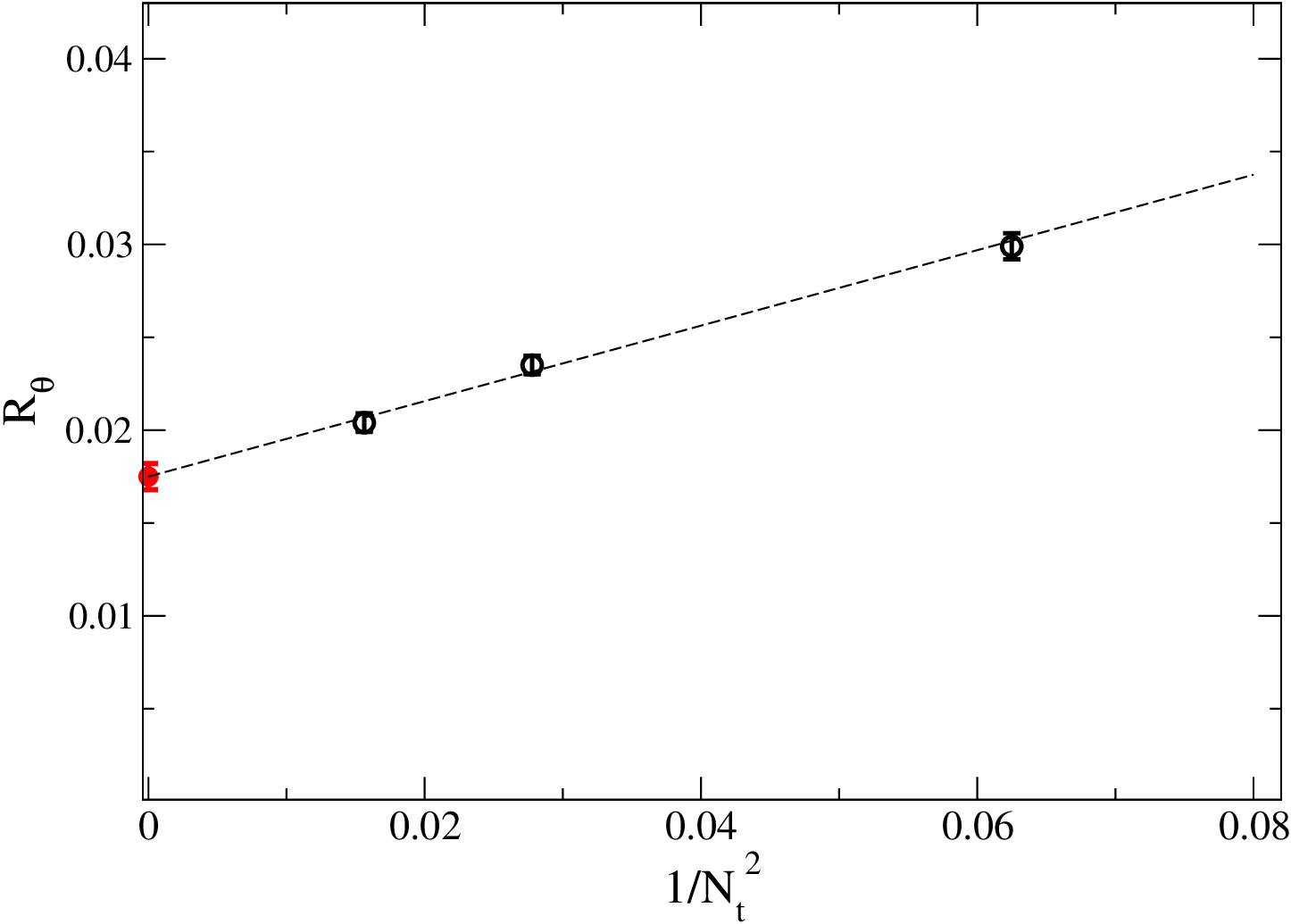}
\caption{$R_\theta$ as a function of $1/N_t^2$. The point at $1/N_t = 0$ is the continuum limit
extrapolation, assuming
$O(a^2)$ corrections.}
\label{fig:contlim}
\end{figure}

Assuming 
$O(a^2)$ corrections we can extrapolate
the continuum value $R_\theta =  0.0175(7)$,
$\chi^2/{\rm d.o.f.} \simeq 0.97$ 
(see Fig.~\ref{fig:contlim}).
Our result is therefore that $T_c$ decreases in presence of a real
non-zero $\theta$ parameter. This is in agreement with the large $N$ expectation
that we discuss in the following, as well as with arguments based on the semi-classical approximation
discussed in Ref.~\cite{unsal} for $N = 2$ and with model computations~\cite{kouno}.

{\it Large $N$ estimate} -- We present now a simple argument to estimate the 
dependence of $T_c$ on $\theta$ in the large $N$
limit.
Since the transition is first order, around the critical temperature
we can define two different free energy densities, $f_c(T)$ and
$f_d(T)$, corresponding to the two different phases, 
confined and deconfined,
which cross each other at $T_c$ with 
two different slopes. The slope difference is related to the 
latent heat. Indeed the energy density is
\beq
\epsilon = \frac{T^2}{V_s} \frac{\partial}{\partial T} \log Z \,\, ; \,\,\,\,\,\,\,
Z = \exp \left( -\frac{V_s f(T)}{T} \right) \, 
\eeq
hence 
$\epsilon = -T^2\ \partial (f/T) / \partial T$.
Close enough to a first order transition we may assume, apart from constant
terms, $f_c/T = A_c t + O(t^2)$ and $f_d/T = A_d t + O(t^2)$, where
$t \equiv (T - T_c)/T_c$ is the reduced temperature. The latent heat
is therefore $\Delta \epsilon = \epsilon_d - \epsilon_c = T_c (A_c - A_d)$.

A non-zero $\theta$ modifies the free energy, 
at the lowest order,
as follows:
\beq
{f(T,\theta)} = {f(T,\theta = 0)} +
\chi(T)\, {\theta^2}/{2}\ + O(\theta^4)
\eeq
where $\chi(T)$ is the topological susceptibility. $\chi(T)$ is 
in general different in the two phases,
dropping at deconfinement~\cite{susc_ft,lucini_1,vicari_ft}, hence the condition for 
free energy equilibrium, $f_c = f_d$, which gives the value of $T_c$, will 
change as a function of $\theta$. The dependence of $\chi$ on $T$ simplifies
in the large $N$ limit, being independent of $T$ in
the confined phase and vanishing in the 
deconfined one~\cite{lucini_1,vicari_ft}.
Hence we can write, for $N \to \infty$,
\beq
f_c/T \simeq A_c t + (\chi/T)\ \theta^2/2 \,\,\, ; \,\,\,\,\,\,\,\,\,\,\, 
f_d/T \simeq A_d t 
\eeq
where $\chi$ is, from now on, the $T = 0$ topological susceptibility.
The equilibrium condition then reads
$(A_c - A_d)\ t = (\chi/T_c) \theta^2/2 + O(\theta^4)$, giving
\beq
\frac{T_c (\theta)}{T_c (0)} = 
1 - \frac{\chi}{2 \Delta \epsilon} \theta^2 + O(\theta^4)  
\label{prediction}
\eeq  
In the large $N$ limit we have~\cite{vicari_rep,lucini_1,lucini_2}, 
$$
\frac{\chi}{\sigma^2} \simeq 0.0221(14) \, ; \,\,
\frac{\Delta \epsilon}{N^2 T_c^4} \simeq 0.344(72) \, ; \,\,
\frac{T_c}{\sqrt{\sigma}} \simeq 0.5970(38)
$$
apart from $1/N^2$ corrections, hence we get
\beq
R_\theta = \frac{\chi}{2 \Delta \epsilon} \simeq
\frac{0.253(56)}{N^2} + O(1/N^4) \; .
\label{largenpred}
\eeq

The leading $1/N$ estimate for $SU(3)$ is then
$R_\theta \simeq 0.0281(62)$. This is larger 
than our determination, even
if marginally compatible with it: 
a possible interpretation is that for $SU(3)$ the behavior
of $\chi$ at $T_c$ is smoother than the sharp drop
to zero that we have assumed.

Notice that the $1/N^2$ dependence of $R_\theta$ is in agreement with 
general arguments~\cite{witten} predicting the free energy to be 
a function of the variable $\theta/N$ as $N \to \infty$
(see also Refs.~\cite{vicari_rep,unsal}). For the same reason
we expect $O(\theta^4)$ corrections to Eq.~(\ref{prediction})
to be of $O(1/N^4)$: they are indeed related to $O(\theta^4)$ corrections to the 
free energy, which have been measured at $T = 0$ by lattice 
simulations~\cite{vicari_b4,nostro_b4,giusti} and are known to be 
small and of order $1/N^2$.

It would be interesting to extend the present study to 
$N > 3$, in order to check the prediction in Eq.~(\ref{largenpred}), 
and to $N = 2$, in order to compare with the results of Ref.~\cite{unsal}.
\\

We conclude with a few remarks and speculations regarding the  
phase structure in the $T-\theta^2$ plane. In Fig.~\ref{fig:tctheta}
we have drawn the critical line, for different $N_t$ and 
up to $\theta^2$ terms, as fitted
from $\theta^2 < 0$ simulations, and its continuation to $\theta^2 > 0$; 
however other transition lines may be present, 
as it happens for the $T-\mu_B^2$ plane. 
For $\mu_B^2 < 0$ one finds unphysical transitions, known
as Roberge-Weiss lines~\cite{rw}, which are linked to the periodicity
of the theory in terms of imaginary $\mu_B$. In the case of a $\theta$ 
parameter, no periodicity is expected for imaginary $\theta$,
CP inviariance being explicitely broken for any $\theta_I \neq 0$,
hence we cannot predict
other possible transitions for $\theta^2 <0$.
A $2 \pi$ periodicity is instead expected for real values of $\theta$,
with the possible presence of a phase transition at $\theta = \pi$
where CP breaks spontaneously.

Our simulations have given evidence, for $\theta^2 < 0$, only
for a deconfinement transition line, describable
by a $\theta^2$ behavior up to $|\theta| \sim \pi$. We expect continuity
of such behavior, at least for small real $\theta$,
 while non-trivial corrections may appear as $\theta$ approaches $\pi$.
However, following Ref.~\cite{witten} and the arguments
above, we speculate that, at least for large $N$, $T_c(\theta)$ be a multibranched
function, dominated by the quadratic term down to $\theta = \pi$
\beq
T_c(\theta)/T_c(0) \simeq 1 - R_\theta 
\min_k\left(\theta+2\pi k\right)^2
\eeq
where $k$ is a relative integer: in this case 
periodicity in $\theta$ implies
cusps for $T_c(\theta)$ at $\theta = (2 k + 1)\, \pi$, where
the deconfinement line could meet the 
CP breaking transition present also at $T = 0$. Therefore
the phase diagram at real $\theta$ 
could have some analogies with that found at imaginary $\mu_B$.

\noindent {\bf Acknowledgements:}
We thank C.~Bonati, A.~Di~Giacomo, B.~Lucini, E.~Shuryak, 
M.~Unsal and E.~Vicari  
for useful discussions. 
We acknowledge the use of the computer facilities of the INFN Bari Computer 
Center for Science, of the INFN-Genova Section and 
of the CSN4 Cluster in Pisa.

\end{document}